\documentclass{acmsiggraph}

\usepackage[normalem]{ulem}

\makeatletter
\newcommand\footnoteref[1]{\protected@xdef\@thefnmark{\ref{#1}}\@footnotemark}
\makeatother

\newcommand{\tagn}[1]{\textsf{#1}}

\usepackage{steinmetz}
\usepackage{physics}
\usepackage{graphicx}
\usepackage{amssymb}
\usepackage[usenames,dvipsnames]{color}
\usepackage{amsmath}
\usepackage{mathtools}
\usepackage{multirow}
\usepackage{algorithm}
\usepackage{algorithmic}
\usepackage{amsfonts}
\usepackage{bbm}
\usepackage[english]{babel} 

\definecolor{meyblue}{rgb}{0.4,0.4,1.0}
\definecolor{meygreen}{rgb}{0.4,1.0,0.4}
\definecolor{meyred}{rgb}{1.0,0.4,0.4}
\definecolor{meygray}{rgb}{0.6,0.6,0.6}
\definecolor{meyorange2}{rgb}{0.8,0.5,0.1}
\definecolor{meyblue2}{rgb}{0.4,0.4,0.8}

\newcommand{\aaron}[1]{\textcolor{blue}{\bf [Aaron: #1] }}
\newcommand{\ersin}[1]{\textcolor{brown}{\bf [Ersin: #1]}}

\newcommand{\mypara}[1]{\paragraph*{#1}}

\def\edit#1{{#1}} 
\def\edittwo#1{{#1}} 

\graphicspath{{figures/}}

\newcommand{\bc}{\mathbf{c}}

\newcommand{\bL}{\mathbf{L}}

\newcommand{\bM}{\mathbf{M}}

\newcommand{\bp}{\mathbf{p}}

\newcommand{\bx}{\mathbf{x}}
\newcommand{\bw}{\mathbf{w}}

\newcommand{\by}{\mathbf{y}}

\newcommand{\cD}{\mathcal{D}}

\newcommand{\authspace}[0]{\hspace{.2cm}}
\newcommand{\affspace}[0]{\hspace{.2cm}}

\newcommand\printthanks{%
  \renewcommand{\thefootnote}{}%
  \footnotetext[0]{%
\hspace*{-1.8em}%
  Li Yi co-developed and implemented the method; the other authors are in alphabetical order.This work started during Li Yi's internship at Adobe Research. This work is supported by NSF grants DMS-1521608 and DMS-1546206, ONR grant MURI N00014-13-1-0341, and Adobe Systems Inc.
}%
  \renewcommand{\thefootnote}{\arabic{footnote}}
}

\title{Learning Hierarchical Shape Segmentation and Labeling\\from Online Repositories}

\author{
\hspace{-1cm}
Li Yi\textsuperscript{1}\authspace
Leonidas Guibas\textsuperscript{1} \authspace
Aaron Hertzmann\textsuperscript{2} \authspace
Vladimir G. Kim\textsuperscript{2} \authspace
Hao Su\textsuperscript{1} \authspace
Ersin Yumer\textsuperscript{2} \authspace
\\
\\
\hspace{-1cm}
\textsuperscript{1}Stanford University \affspace \textsuperscript{2}Adobe Research
}

\pdfauthor{}

\TOGonlineid{0313}


\CopyrightYear{2017} 
\setcopyright{acmcopyright} 
\conferenceinfo{SIGGRAPH '17 Technical Paper,}{July 30-August 03, 2017, Los Angeles, CA, USA}
\isbn{978-1-4503-4997-0/17/07}\acmPrice{\$15.00}
\doi{http://dx.doi.org/10.1145/3072959.3073652}

\begin{document}


\teaser{
  \includegraphics[width=1\textwidth]{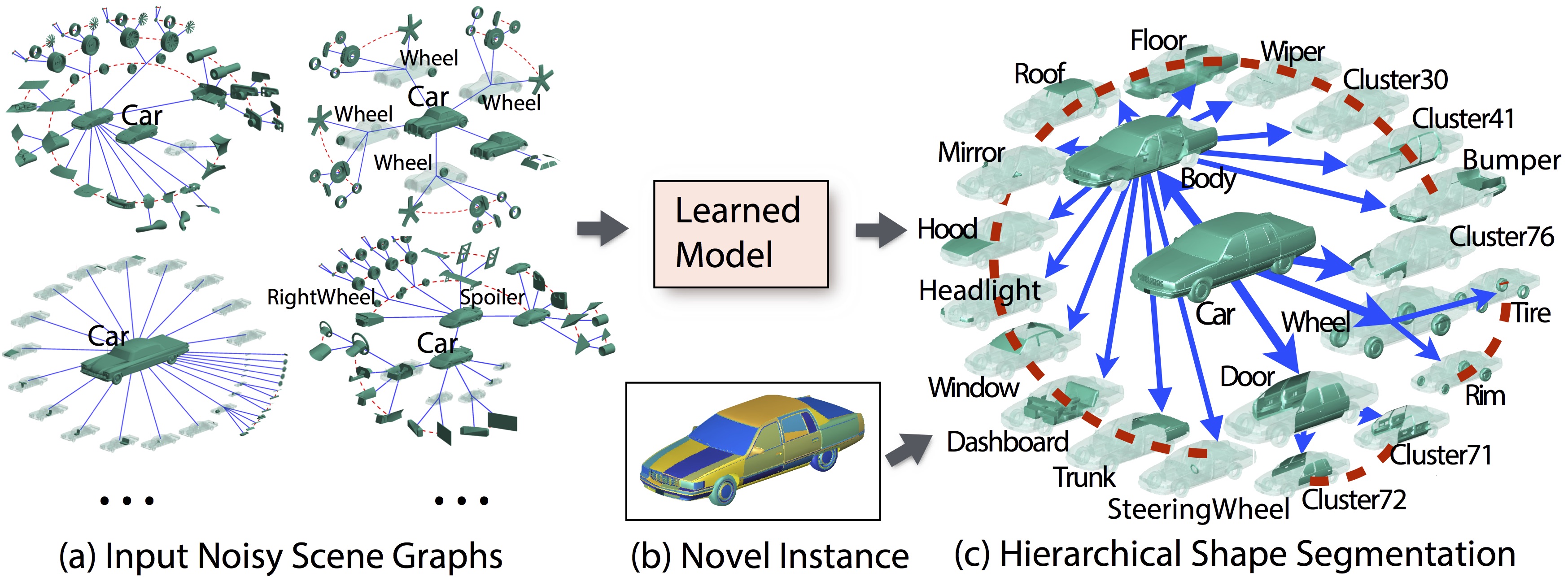}
  \vspace{-0.5cm}
  \caption{ \label{fig:teaser} 
Large online model repositories contain abundant additional data beyond 3D geometry, such as part labels and artist's part decompositions, flat or hierarchical.
We tap into this trove of sparse and noisy noisy data to train a network for simultaneous hierarchical shape structure decomposition and labeling.  Our method learns to take new geometry, and segment it into parts, label the parts, and place them in a hierarchy.
In this paper, we visualize scene graphs with a circular visualization, in which the root node is near the center. Blue lines indicate parent-child relationships, and red dashed arcs connect siblings.  The input geometry in online databases are broken as connected components, visualized in the input as random colors.
}
}

\maketitle

\begin{abstract}
We propose a method for converting geometric shapes into hierarchically segmented parts with part labels. Our key idea is to train category-specific models from the scene graphs and part names that accompany 3D shapes in public repositories. These freely-available annotations represent an enormous, untapped source of information on geometry. However, because the models and corresponding scene graphs are created by a wide range of modelers with different levels of expertise, modeling tools, and objectives, these models have very inconsistent segmentations and hierarchies with sparse and noisy textual tags. Our method involves two analysis steps. First, we perform a joint optimization to simultaneously cluster and label parts in the database while also inferring a canonical tag dictionary and  part hierarchy. We then use this labeled data to train a method for hierarchical segmentation and labeling of new 3D shapes. We demonstrate that our method can mine complex information, detecting hierarchies in man-made objects and their constituent parts, obtaining finer scale details than existing alternatives. We also show that, by performing domain transfer using a few supervised examples, our technique outperforms fully-supervised  techniques that require hundreds of manually-labeled models.


\end{abstract}

%
%


%
%



\keywordlist

\conceptlist

\printthanks

\section{Introduction}


Segmentation and labeling of 3D shapes is an important problem in geometry processing.  These structural annotations are critical for many applications, such as animation, geometric modeling, manufacturing, and search~\cite{mitra2013structure}.  Recent methods have shown that, by supervised training from labeled shape databases, state-of-the-art performance can be achieved on mesh segmentation and part labeling \cite{kalogerakis2010learning,yi2016scalable}.  However, such methods rely on carefully-annotated databases of shape segmentations, which is an extremely labor-intensive process. Moreover, these methods have used coarse segmentations into just a few parts each, and do not capture the fine-grained, hierarchical structure of many real-world objects. 
Capturing fine-scale part structure is very difficult with non-expert manual annotation; it 
\edit{is difficult even to determine the set of parts and labels to separate.}
%
%
Another option is to use unsupervised methods that work without annotations by analyzing  geometric patterns~\cite{van2013co}. Unfortunately, these methods do not have access to the full semantics of shapes and as a result often do not identity parts that are meaningful to humans, nor can they apply language labels to models or their parts. Additionally, typical co-analysis techniques do not easily scale to large datasets.



We observe that, when creating 3D shapes, artists often provide a considerable amount of extra structure with the model. In particular, they separate parts into hierarchies represented as \emph{scene graphs}, and annotate individual parts with textual names. In surveying the online geometry repositories, we find that most shapes are provided with these kinds of user annotations. Furthermore, there are often thousands of models per category available to train from. Hence, we ask: can we exploit this abundant and freely-available metadata to analyze and annotate new geometry?


Using these user-provided annotations comes with many challenges. 
\edit{For instance, Figure \ref{fig:teaser}(a) shows four typical scene graphs in the \textsf{car} category, created by four different authors.  Each one has a different part hierarchy and set of parts, e.g., only two of the scene graphs have the steering wheel of the car as a separate node.
The hierarchies have different depths; some are nearly-flat hierarchies and some are more complex. Only a few parts are given names in each model.} 
\edit{ Despite this variability, inspecting these models reveal common trends, such as certain parts that are frequently segmented, parts that are frequently given consistent names, and pairs of parts that frequently occur in parent-child relationships with each other. For example, the tire is often a separate part, it is usually the child of the wheel, and usually has a name like \textsf{tire} or \textsf{RightTire}.}
 \edit{Our goal is to exploit these trends, while being robust to the many variations in names and hierarchies that different model creators use.}

This paper proposes to learn shape analysis from these messy, user-created datasets,
thus leveraging the freely-available annotations provided by modelers. Our main goal is to automatically discover common trends in part segmentation, labeling, and hierarchy. Once learned, our method can be applied to new shapes that consist of geometry alone: the new shape is automatically segmented into parts, which are labeled and placed in a hierarchy. Our method can also be used to clean-up the existing databases. Our method is designed to work with large training sets, learning from thousands of models in a category.
Because the annotations are uncurated, sparse (within each shape) and irregular, this problem is an instance of weakly-supervised learning.  

Our approach handles each shape category (e.g., cars, airplanes, etc.) in a dataset separately. For a given shape category, we first identify the commonly-occurring part names within that class, and manually condense this set, combining synonyms, and removing uninformative names.  We then perform an optimization that simultaneously (a) learns a metric for classifying parts, \edit{(b) assigns names to unnamed parts where possible, (c) clusters other unnamed parts,} (d) learns a canonical hierarchy for parts in the class, and (e) provides a consistent labeling to all parts in the database. 
Given this annotation of the training data, we then learn to hierarchically segment new models, using a Markov Random Field (MRF) segmentation algorithm.   Our algorithms are designed to scale to training on large datasets by mini-batch processing.
%
We use these outputs to train a hierarchical segmentation model. Then, given a new, unsegmented mesh, we can apply this learned model to segment the mesh, transfer the tags, and infer the part hierarchy.

We use our method to analyze shapes from ShapeNet \cite{chang2015shapenet}, a large-scale dataset of 3D models and part graphs obtained from online repositories. We demonstrate that our method can mine complex information detecting hierarchies in man-made objects and their constituent parts, obtaining finer scale details than existing alternatives. While our problem is different from what has been explored in previous research, we perform two types of quantitative evaluations. First, we evaluate different variants of our method by holding some tags out, and show that all terms in our objective function are important to obtain the final result. Second, we show that supervised learning techniques require hundreds of manually labeled models until they reach the quality of segmentation that we get without any explicit supervision. ~\edit{We publicly share our code and the processed datasets in order to encourage further research.\footnote{\url{http://cs.stanford.edu/~ericyi/project_page/hier_seg/index.html}}}

\section{Related Work}\label{sec:related}

Recent shape analysis techniques focus on extracting structure from large collections of 3D models~\cite{Xu16}. In this section we discuss recent work on detecting labeled parts and hierarchies in shape collections. 

\mypara{Shape Segmentation and Labeling.}
Given a sufficient number of training examples, it is possible to learn to segment and label novel geometries~\cite{kalogerakis2010learning,yumer2014co,Guo:2015}. While supervised techniques achieve impressive accuracy, they require dense training data for each new shape category, which significantly limits their applicability. To decrease the cost of data collection, researchers have developed methods that rely on crowdsourcing~\cite{Chen:2009}, active learning~\cite{Wang12}, or both~\cite{yi2016scalable}. However, this only decreases the cost of data collection, but does not eliminate it. Moreover, these methods have not demonstrated the ability to identify fine-grained model structure, or hierarchies. One can rely solely on consistency in part geometry to extract meaningful segments without  supervision~\cite{Golovinskiy09,Sidi11,Huang11,Hu12,kim2013learning,Huang14}. However, since these methods do not take any human input into account, they typically only detect coarse parts, and do not discover semantically salient regions where geometric cues fail to encapsulate the necessary discriminative information. 

In contrast, we use the part graphs that accompany 3D models to weakly supervise the shape segmentation and labeling. This is similar in spirit to existing unsupervised approaches, but it mines semantic guidance from ambient data that accompanies most available 3D models. 

Our method is an instance of weakly-supervised learning from data on the web. A number of related problems have been explored in computer vision, including learning classifiers and captions from user-provided images on the web, e.g., \cite{izadinia2015,LiSurvey,ordonez2011}, or image searches, e.g., \cite{chenGuptaWebly}.

\mypara{Shape Hierarchies.} 
Previous work attempted to infer scene graphs based on symmetry~\cite{wang_eg11} or geometric matching~\cite{van2013co}. However, as with unsupervised segmentation techniques, these methods only succeed in a presence of strong geometric cues. To address this limitation, Liu et al.~\shortcite{Liu14} proposed a method that learns a probabilistic grammar from examples, and then uses it to create consistent scene graphs for unlabeled input. However, their method requires accurately labeled example scene graphs.   Fisher et al.~\shortcite{fisher2011characterizing} use scene graphs from online repositories, focusing on arrangements of objects in scenes, whereas we focus on fine-scale analysis of individual shapes.

In contrast, we leverage the scene graphs that exist for most shapes created by humans. Even though these scene graphs are noisy and contain few meaningful node names (Figure \ref{fig:teaser}(a)), we show that it is possible to learn a consistent hierarchy by combining cues from corresponding sparse labels and similar geometric entities in a joint framework. Such label correspondences not only help our clusters be semantically meaningful, but also help us discover additional common nodes in the hierarchy.


\section{Overview}\label{sec:overview}


Our goal is to learn an algorithm that, given a shape from a specific class \edit{(e.g., cars or airplanes)}, can segment the shape, label the parts, and place the parts into a hierarchy.
Our approach is to train based on geometry downloaded from online model repositories.
\edit{Each shape is composed of 3D geometry segmented into distinct parts; each part has an optional textual name, and the parts are placed in a hierarchy. The hierarchy for a single model is called a scene graph. As discussed above, different training models may be segmented in different hierarchies; our goal is to learn from trends in the data as to which parts are often segmented, how they are typically labeled, and which parts are typically children of other parts.}

%
%

We break the analysis into two sub-tasks:
\begin{itemize}
\item \textbf{Part-Based Analysis} (Section \ref{sec:method}). 
\edit{Given a set of meshes in a specific category  and their original messy scene graphs,} we identify the dictionary of distinct parts for a category, and place them into a canonical hierarchy. This dictionary includes both parts with user-provided names (e.g., \texttt{wheel}) and a clustering of unnamed parts.  All parts on the training meshes are labeled according to the part dictionary. 

\item  \textbf{Hierarchical Mesh Segmentation} (Section \ref{sec:seg}).  We train a method to segment a new mesh into a hierarchical segmentation, using the labels and hierarchy provided by the previous step. For parts with textual names, these labels are also transferred to the new parts.  
\end{itemize}

We evaluate with testing on hold-out data, and qualitative evaluation. In addition, we show how to adapt our model to a benchmark dataset.

\begin{figure}[t!]
	\centering
	\includegraphics[width=1\columnwidth]{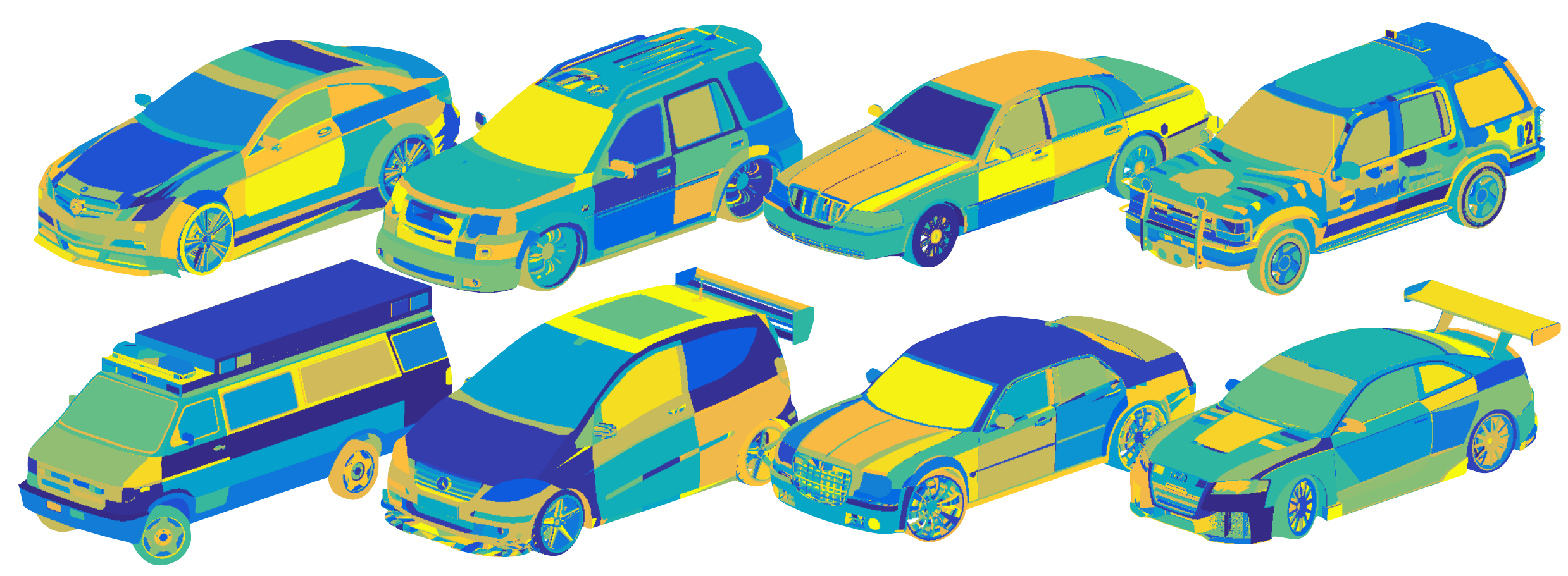}
	\caption{A visualization of connected components in ShapeNet cars,  illustrating that each connected component is usually a sub-region of a single part.}
	\label{fig:viscc}
\end{figure}

Our method makes two additional assumptions. 
First, our feature vector representations assume consistently-oriented meshes, 
following the representation in \edittwo{ShapeNetCore} \cite{chang2015shapenet}. 
Second, the canonical hierarchy requires that every type of part has only one possible parent label, e.g., our algorithm might infer that the parent of a \texttt{headlight} is always the \texttt{body}, if this is frequently the case in the training data.

In our segmentation algorithm, we usually assume that each connected component in the mesh belongs to a single part.
This can be viewed as a form of over-segmentation assumption (e.g.,~\cite{van2013co}), and we found it to be generally true for our input data, e.g., see Figure~\ref{fig:teaser}(b) and~\ref{fig:viscc}. We  show results both with and without this assumption in Section \ref{sec:results} and in the Supplemental Material.

\section{Part-Based Analysis}\label{sec:method}

The first step of our process takes the shapes in one category as input, and identifies a dictionary of parts for that category, a canonical hierarchy for the parts, and a labeling of the training meshes according to this part dictionary.  Each input shape $i$ is represented by a scene graph: a rooted directed tree $H_i = \{X_i, E_i\}$, where nodes are parts with geometric features $X_i = \{\bx_{ij} | j=1,...,|X_i|\}$ and each edge $(j,k)\in E_i$ indicates that part $(i,k)$  is a child of part $(i,j)$. We manually pre-process the user-provided part names into a tag dictionary $T$, which is a list of part names relevant for the input category (Table \ref{tab:labelfrequency}).~\edit{One could imagine discovering these names automatically. We opted for the manual processing, since the vocabulary of words that appear in ShapeNet part labels is fairly limited, and there are many irregularities in the label usage, e.g., synonyms and mispellings.}
The parts with a label from the dictionary are then assigned corresponding tags $t_{ij}$. Note that many parts are untagged, either because no names were provided with the model, or the user-provided names did not map onto names in the dictionary.    Note also that $j$ is indexes parts within a shape independent of tags; e.g., there is no necessary relation between $(i,j)$ and part $(k,j)$.
Each graph has a root node, which has a special root tag, and no parent.
For non-leaf nodes, the geometry of any node is the union of geometries of its children.

To produce a dictionary of parts, we could directly use the user-provided tags, and then cluster the untagged parts. However, this naive approach would have several intertwined problems. First, the user-provided tags may be incorrect in various ways: missing tags for known parts (e.g., a wheel not tagged at all), tags given only at a high-level of the hierarchy (e.g., the rim and the tire are not segmented from the wheel, and they are all tagged as \tagn{wheel}), and tags that are simply wrong. 
The clustering itself depends on a distance metric, which must be learned from labels.
We would like to have tags be applied as broadly and accurately as possible, to provide as much clean training data as possible for labeling and clustering, and to correctly transfer tags when possible.
Finally, we would also like to use a parent-child relationships to constrain the part labeling (so that a \tagn{wheel} is not the child of a \tagn{door}), but plausible parent-child relationships are not known a priori.

We address these problems by jointly optimizing for all unknowns: the distance metric, a dictionary of $D$ parts, 
a labeling of parts according to this dictionary, and a probability distribution over parent-child relationships.  The labeling of model parts is also done probabilistically, by the \edittwo{Expectation-Maximization (EM)} algorithm \cite{neal1998view}, where the hidden variables are the part labels.  The distance metric is encoded in a embedding function $f: \mathbb{R}^S \rightarrow \mathbb{R}^F$, which maps a part represented by a shape descriptor $\bx$ (Appendix~\ref{appx:features}) to a lower-dimensional feature space. The function $f$ is represented as a neural network (Figure~\ref{fig:network_arc}).  
Each canonical part $k$ has a representative cluster center $\bc_k$ in the feature space, so that a new part can be classified by nearest-neighbors distance in the feature space. Note that the clusters do not have an explicit association with tags: our energy function only encourages parts with the same tag to fall in the same cluster. As a post-process, we match tag names to clusters where possible.

We model parent-child relationships with a matrix $\bM \in \mathbb{R}^{D \times D}$, where $\bM_{uv}$ is, for a part in cluster $v$, the probability that its parent has label $u$. After the learning stage, $\bM$ is converted to a deterministic canonical hierarchy over all of the parts.

Our method is inspired in part by the semi-supervised clustering method of Basu et al.~\shortcite{basu2004probabilistic}. In contrast to their linear embedding of initial features for metric learning, we incorporate a neural network embedding procedure to allow non-linear embedding in the presence of constraints, and use an EM soft clustering. In addition, Basu et al.~\shortcite{basu2004probabilistic} do not take hierarchical representations into consideration, whereas our data is inherently a hierarchical part tree.

\subsection{Objective function}

The EM objective function is:
\begin{equation}
E(\theta, \bp,\bc,\bM) = \lambda_c E_c + \lambda_s E_s + \lambda_d E_d + \lambda_m E_m - H
\label{eq:EMobj}
\end{equation}
where $\theta$ are the parameters of the embedding $f$, $\bp$ are the label probabilities such that $p_{ijk}$ represents the probability of the $j^\textrm{th}$ part of $i^\textrm{th}$ shape be assigned to $k^\textrm{th}$ label cluster, and $\bc_{1:D}$ are the unknown cluster centers.
We set $\lambda_c=0.1, \lambda_s=1, \lambda_d=1, \lambda_m=0.05$ throughout all experiments.

The first two terms, $E_c$ and $E_s$, encourage the concentration of clusters in the embedding space; $E_d$ encourages the separation of visually dissimilar parts in embedding space; $E_m$ is introduced to estimate the parent-child relationship matrix $\bM$; the entropy term $H = -\sum p \ln p$ is a consequence of the derivation of the EM objective (Appendix \ref{app:EM}) and is required for correct estimation of probabilities. We next describe the energy terms one by one.

Our first term favors part embeddings to be near their corresponding cluster centroids:
\begin{equation}
E_c = \sum_{ (i,j) }\sum_{k \in 1:D} p_{ijk} ||f(\bx_{ij})-\bc_k||_1 
\label{eq:Ec}
\end{equation}
where $f$ is the embedding function $f(\cdot)$, represented as a neural network and parametrized by a vector $\theta$.
The network is described in Appendix \ref{appx:features}.

Second, our objective function constrains the embedding, by favoring small distances for parts that share the same input tag, and for parts that have very similar geometry:
\begin{align}
\label{eq:Es}
E_s &= \sum_{(x_{ij},x_{kl})\in\mathbb{S}} ||f(\bx_{ij})-f(\bx_{kl})||_1\text{, where} \\
&(x_{ij},x_{kl})\in\mathbb{S} \text{ iff }  t_{ij} = t_{kl} \text{ or } \|\bx_{ij}-\bx_{kl}\|_2^2 \leq \delta
\nonumber
\end{align}
%
We extract all tagged parts and sample pairs from them for the constraint. 
We set $\delta=0.1$ to a small constant to account for near-perfect repetitions of parts, and ensure that these parts are assigned to the same cluster. 

Third, our objective favors separation in the embedded space by a margin $\sigma_d$ between parts on the same shape that are not expected to have the same label: 
\begin{align}
\label{eq:Ed}
E_d &= \underset{(x_{ij},x_{il})\in\mathbb{D}}{\sum}\text{max}(0,\sigma_d-||f(\bx_{ij})-f(\bx_{il})||_1)\text{, where} \\
&(x_{ij},x_{il})\in\mathbb{D} \text{ if } (x_{ij},x_{il}) \not\in \mathbb{S}. \nonumber
\end{align}
We only use parts from the same shape in $\mathbb{D}$, since we believe it is generally reasonable to assume that parts on the same shape with distinct tags or distinct geometry have distinct labels.

Finally, we score the labels of parent-child pairs by how well they match the overall parent-child label statistics in the data, using the negative log-likelhood of a multinomial:
\begin{eqnarray}
\label{eq:Em}
E_m = -\underset{\ell_1, \ell_2 \in 1:D\times 1:D}{\sum} 
~~
\underset{i}{\sum}
~~
  \underset{(j,k) \in E_i}{\sum} 
  p_{i j \ell_1} p_{i k \ell_2}\ln \bM_{\ell_1 \ell_2}
\end{eqnarray}

\subsection{Generalized EM algorithm}

We optimize the  objective function (Equation~\ref{eq:EMobj}) by alternating between E and M steps. We solve for the soft labeling $\bp$ in the E-step, and the other parameters, $\Theta=\{\theta, \bc, \bM\}$, in the M-step, where $\theta$ are the parameters of the embedding $f$. 

\mypara{E-step.} Holding the model parameters $\Theta$ fixed, we optimize for the label probabilities $\bp$:
\begin{equation}
\begin{aligned}
&\underset{\bp}{\text{minimize}~}
\lambda_c E_c + \lambda_m E_m - H
\end{aligned}
\label{eqn:clustering}
\end{equation}
We optimize this via coordinate descent, by iterating $5$ times over all coordinates. \edit{The update is given in Appendix \ref{app:pupdate}.}

\mypara{M-step.} Next, we \edit{hold the soft clustering $\bp$ fixed} and optimize the model parameters $\Theta$ by solving the following subproblem:
\begin{equation}
\begin{aligned}
& \underset{\Theta}{\text{minimize}\!}
& & \lambda_c E_c + \lambda_s E_s + \lambda_d E_d + \lambda_m E_m
\end{aligned}
\label{eqn:embedding}
\end{equation}
\edit{We use stochastic gradient descent updates for $\theta$ and $\bc_{1:D}$, as is standard for neural networks, while keeping $\bp, \bM$ fixed. 
The parent-child probabilities $\bM$ are then computed as:
\begin{align}
\bM &\leftarrow \mathrm{normc}\left(\sum_i \sum_{(j,k)\in E_i} \bp_{ij} \bp^T_{ik}\right) + \epsilon
\label{eqn:optm}
\end{align}
}
where $\mathrm{normc}(\cdot)$ is a column-wise normalization function to guarantee $\sum_i\bM_{ij}=1$. $\bp_{ij}$ and $\bp_{ik}$ are the cluster probability vectors that correspond to parts $x_{ij}$ and $x_{ik}$ of the same shape, respectively. $\epsilon=1\times10^{-6}$ in our experiments, to prevent cluster centers from stalling at zero.
\edit{Since each column of $\bM$ is a separate multinomial distribution, the update in Eq.~\ref{eqn:optm} is the standard multinomial estimator.}

\mypara{Mini-batch training.} 
The dataset for any category is far too large to fit in memory, and so, in practice, we break the learning process into mini-batches. Each mini-batch includes 50 geometric models at a time. For the set $\mathbb{S}$, 20,000 random pairs of parts are sampled across models in the mini-batch. 30 epochs (passes over the whole dataset) are used.

For each mini-batch, the E-step is computed as above.
In the mini-batch M-step, the embedding parameters $\theta$ \edit{and cluster centers $\bc$} are updated by standard stochastic gradient descent (SGD) updates, using Adam updates~\cite{adam}. For the  hierarchy $\bM$, we use Stochastic EM updates~\cite{Cappe:2009}, which are more stable and efficient than gradient updates. 
The sufficient statistics are computed for the minibatch:
\begin{align}
\bar{\bM}_{\mathrm{mb}} &= \sum_i \sum_{(j,k)\in E_i} \bp_i \bp_j^T
\label{eqn:updateM-minibatch}
\end{align}
Running averages for the sufficient statistics 
are updated after each mini-batch:
\edit{
\begin{equation}
\bar{\bM} \leftarrow(1-\eta)\bar{M} + \eta \bar{M}_{\mathrm{mb}}
\label{eqn:updatecm_minibatch}
\end{equation}
}
where $\eta=0.5$ in our experiments.
Then, the estimates for  $\bM$ are computed from the current sufficient statistics by:
\begin{align}
\bM & \leftarrow \mathrm{normc}(\bar{\bM}) + \epsilon
\end{align}

\mypara{Initialization.}
Our  objective, like many EM 
algorithms, requires good initialization. We first initialize the neural network embedding $f(\cdot)$ with  normalized initialization \cite{Glorot10understandingthe}.
For each named tag $t_i$, we specify an initial cluster center $\bc_i$  as the average of the embeddings of all the parts with that tag. The remaining $D$ cluster centroids $\bc_{|T|+1:D}$ are randomly sampled from a normal distribution in the embedding space. 
The cluster label probabiilities $\bp$ are initialized by a nearest-neighbor hard-clustering, and then $\bM$ is initialized by Eq.~\ref{eqn:optm}.

\subsection{Outputs}

\begin{figure}
	\centering
	\includegraphics[width=1\columnwidth]{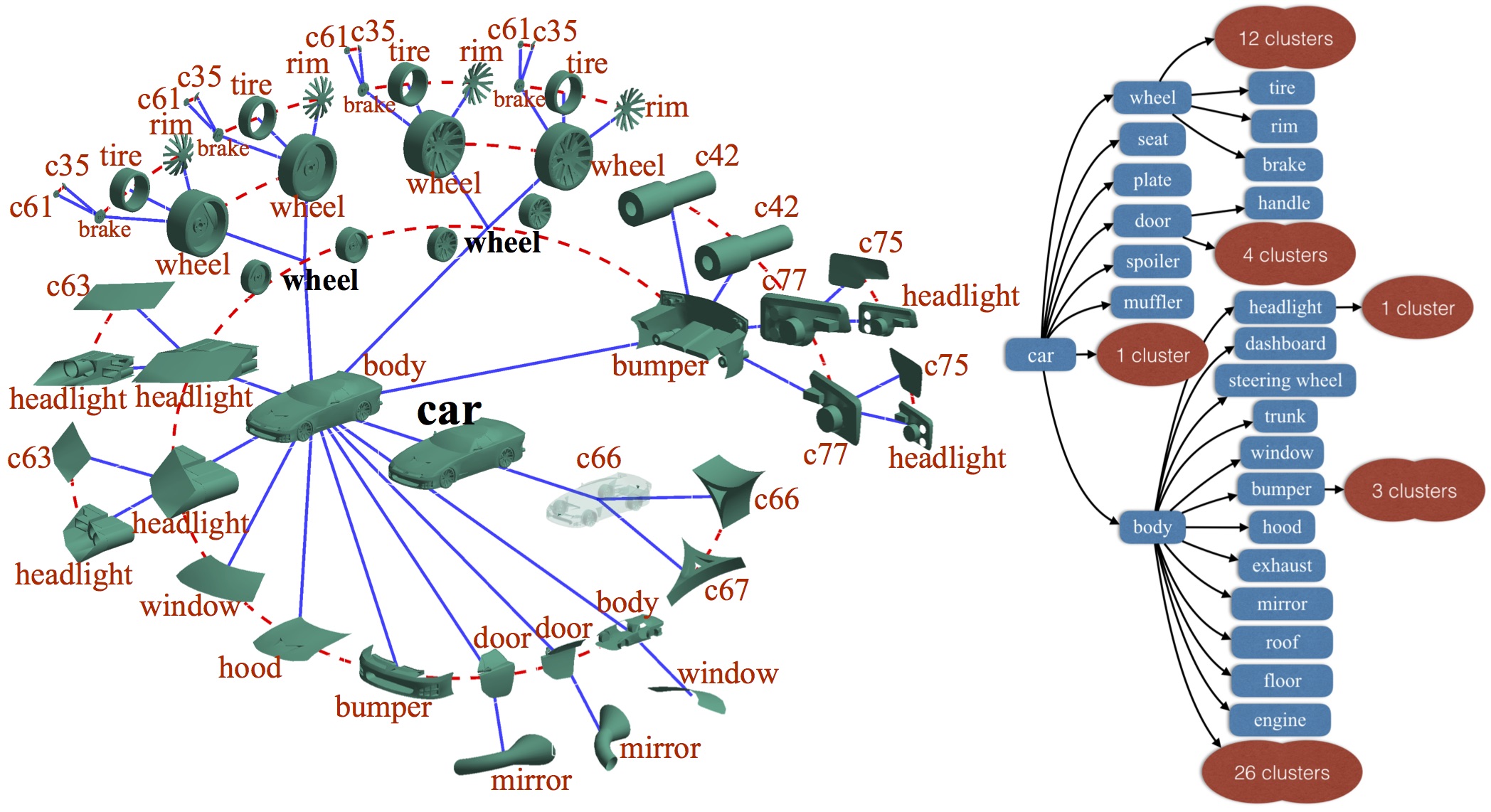}
	\caption{Typical output of our part based analysis. Left: Part labeling for a training shape. Black labels indicate labels given with the input, and red labels were applied by our algorithm.
	Right: Canonical hierarchy. Generic cluster labels indicate newly discovered parts.  Multiple generic clusters are grouped in the visualization, for brevity.
	\label{fig:parttagging}}
\end{figure}

Once the optimization is complete, we compute a canonical hierarchy $\mathcal{T}_M$ from $\bM$ by solving a Directed Minimum Spanning Tree problem, with the root constrained to the entire object. Then, we assign tags to parts in the hierarchy by solving a linear assignment problem that maximizes the number of input tags in each cluster that agree with the tag assigned to their cluster.
As a result, some parts in the canonical hierarchy receive textual names from assigned tags. Unmatched clusters are denoted with generic names \texttt{cluster\kern-0.08em\_N}.
We then label each input part $(i,j)$  with its most likely node in $\mathcal{T}_M$ by selecting $\arg \max_k p_{ijk}$. This gives a part labeling of each node in each input scene graph.
An example of the canonical hierarchy with part names, and a labeled shape, is shown in Figure \ref{fig:parttagging}.

This canonical hierarchy, part dictionary, and part labels for the input scene graphs are then used to train the segmentation algorithm as described in the next section.

\section{Hierarchical Mesh Segmentation}\label{sec:seg}

Given the part dictionary, canonical hierarchy, and per-part labels from the previous section, we next learn to hierarchically segment and label new shapes.
We formulate the problem as labeling each mesh face with one of the leaf labels from the canonical hierarchy. Because each part label has only one possible parent, all of a leaf node's ancestors are unambiguous. In other words, once the leaf nodes are specified, it is straightforward to completely convert the shape into a scene graph, with all the nodes in the graph labeled. In order to demonstrate our approach in full generality, we assume the input shape includes only geometry, and no scene graph or part annotations. However, it should be possible to augment our procedure when such information is available.

\subsection{Unary classifier}

\edit{
We begin by describing a technique for training a classifier for individual faces. This classifier can also be used to classify connected components. In the next section, we build an MRF labeler from this.
Our approach is based on the method of Kalogerakis et al.~\shortcite{kalogerakis2010learning}, but generalized to handle missing leaf labels and connected components, and to use neural network classifiers.
}

\edit{The face classifier is formulated as a neural network that takes geometric features of a face as input, and assigns scores to the leaf node labels for the face.}
The feature vector $\by$ for a face consists of several standard geometric features. The neural network specifies a score function $\bw_i^T g(\by)$, where $\bw_i$ is a weight vector for label $L_i$, and $g(\cdot)$ is a sequence of fully-connected layers and non-linear activation units, applied to $\by$. \edit{The score function is normalized by a softmax function to produce an output probability:
\begin{align}
P_{\mathit{face}}(L_i | \by ) &= \frac{ \exp(\bw_i^T g(\by))}{\sum_{j \in \cD} \exp(\bw_j^T g(\by))}
\end{align}
where $\cD$ is the set of possible leaf node labels.
}
See Appendix~\ref{appx:featuresfaces} for details of the feature vector 
and neural network.

%

\edit{
To train this classifier, we can apply the per-part labels from the previous section to the individual faces. However, there is one problem with doing so: many training meshes are not segmented to the finest possible detail. For example, 
a car wheel might not be segmented into tire and rim, or the windows may not be segmented from the body. In this case, the leaf node labels are not given for each face, but only ancestor nodes are known: we do not know which wheel faces are tire faces. In order to handle this, we introduce a probability table $A(a,b)$.
$A(a,b)$ is the probability of a face taking leaf label $a\in \mathcal{D}$ if the deepest label given for this training face is $b\in \mathcal{L}$. 
For example, $A(\mathsf{tire},\mathsf{wheel})$ is the probability that the correct leaf label for a face labeled as a wheel is tire.
To estimate $A(a,b)$, we first compute the unnormalized $\hat{A}(a,b)$ by counting the number of faces assigned to both label $a$ and label $b$, except that $\hat{A}(a,b)=0$ if $b$ is not an ancestor of $a$ in the canonical hierarchy. 
Then $A$ is determined by normalizing the columns to $A$ to sum to 1: $A = \mathrm{normc}(\hat{A})$.
}

\edit{
We then train the classifier by minimizing the following loss function for $\bw_{1:|\mathcal{D}|}$ and $\theta_g$, the parameters of $g(\cdot)$:
\begin{equation}
E(\theta_g,\bw_{1:|\mathcal{D}|})
= -\sum_k \log\left(
\sum_{i \in \mathcal{D}}
       A(i, b_k) P(L_i | \by_k)
\right)
\label{eq:robustsoftmax}
\end{equation}
where $k$ sums over all faces in the training shapes and $b_k$ is the deepest label assigned to face $k$ as discussed above. 
This loss is the negative log-likelihood of the training data, marginalizing over the hidden true leaf label for each training face, generalizing \cite{izadinia2015}.
We use Stochastic Gradient Descent to minimize this objective.
}

We have also observed that meshes in online repositories are comprised of connected components, and these connected components almost always have the same label for the entire component. For most results presented in this paper, we use connected components as the basic labeling units instead of faces, in order to improve results and speed. 
\edit{
We define the connected component classifier  by aggregating the trained face classifier over all the faces $\mathcal{F}$ of the connected component as follows:
\begin{equation}
        P_{\mathit{CC}}(L_i | \mathcal{F})
= \frac{1}{|\mathcal{F}|}\sum_{k\in\mathcal{F}} P_{\mathit{face}}(L_i | \by_k)
\label{eq:segunary}
\end{equation}
}

\subsection{MRF labeler}

Let $\mathcal{D}$ be the set of leaf node of the canonical hierarchy.  In the case of classifying each connected component, we want to specify one leaf node $L_c \in \mathcal{D}$ for each connected component $c$.
We define the MRF over connected component labels as:
\begin{align}
E(L) =&  \sum_c \psi_{\mathrm{unary}}(L_c) + 
\lambda \sum_{u,v\in\mathbb{E}} \psi_{\mathrm{edge}}(L_{u}, L_{v}) 
\label{eqn:mrfseg}
\end{align}
%
with weight $\lambda$ is set by cross-validation \edit{separately for each shape category} and held constant across all experiments. 
The unary term $\psi_{\mathrm{unary}}$ assesses the likelihood of component $c$ having a given leaf label, based on geometric features of the component, \edit{and is given by the classifier:
\begin{equation}
\psi_{\mathrm{unary}}(L) = -\ln P_{\mathit{CC}}(L|\mathcal{F})
\end{equation}
The edge term prefers adjacent components to have the same label. It is defined as $\psi_{\mathrm{edge}}(L_u,L_v)=\mathrm{td}(u,v)$, where
$\mathrm{td}(L_u,L_v)$ is tree distance between labels $L_u$ and $L_v$ in the canonical hierarchy.
This encourages adjacent labels to be as close in the canonical hierarchy as possible.
For example, $\psi$ is 0 when the two labels are the same, whereas $\psi$ is 2 if they are different but share a common parent. 
To generate the edge set $\mathbb{E}$ in \ref{eqn:mrfseg}, we connect $K$ nearest connected components with this edge, where $K=\min(30, \lceil0.01N_\text{cc}\rceil)$ where $N_\text{cc}$ is the number of connected components in the mesh.
}


Once the classifiers and $\lambda$ are trained, the model can be applied to a new mesh as follows.
First, the leaf labels are determined by optimizing Equation \ref{eqn:mrfseg} using the $\alpha$-$\beta$ swap algorithm~\cite{graphcut}.
 Then, the scene graph  is computed by bottom-up grouping. In particular, adjacent components with same leaf label are first grouped together. Then, adjacent groups with the same parent are grouped at the next level of the hierarchy, and so on.
 
 \edit{
 For the case where connected components are not available, the MRF algorithm is applied for each face. The unary term is given by the face classifier $\psi_{\mathrm{unary}}(L) = -\ln P_{\mathit{face}}(L|\mathcal{F})$.  
 We still need to handle the case where the object is not topologically connected, and so the pairwise term $\psi_\text{edge}(L_u,L_v)$ applies to all faces $u$ and $v$ whose centroids fall into the $K$-nearest neighborhood of each other, and is given by:
\begin{equation}
    \psi_\text{edge}(L_u,L_v)=\lambda_t  \exp( -\kappa_1\frac{2\phi_{u,v}}{\pi}-\frac{d_{u,v}^2}{2(\kappa_2d_r)^2} )\ \mathrm{td}(L_u,L_v)
\end{equation}
where $\phi_{u,v}$ is the angle between the faces,  $d_{u,v}$ is the distance between the face centroids,
$d_r$ is the average distance between a face's centroid and it's nearest face's centroid, and $\kappa_1=5, \kappa_2=2.5$ in all our experiments. $\lambda_t$ is a scale factor to promote faces sharing an edge: $\lambda_t = \left\{\begin{array}{l}10 \quad \text{ if faces $(u,v)$ share an edge,}\\1\phantom{0} \quad \text{ otherwise.}\end{array}\right.$ 
}

\section{Results}
\label{sec:experiments}
\label{sec:results}

\begin{figure}[t!]
	\centering
	\includegraphics[width=1\columnwidth]{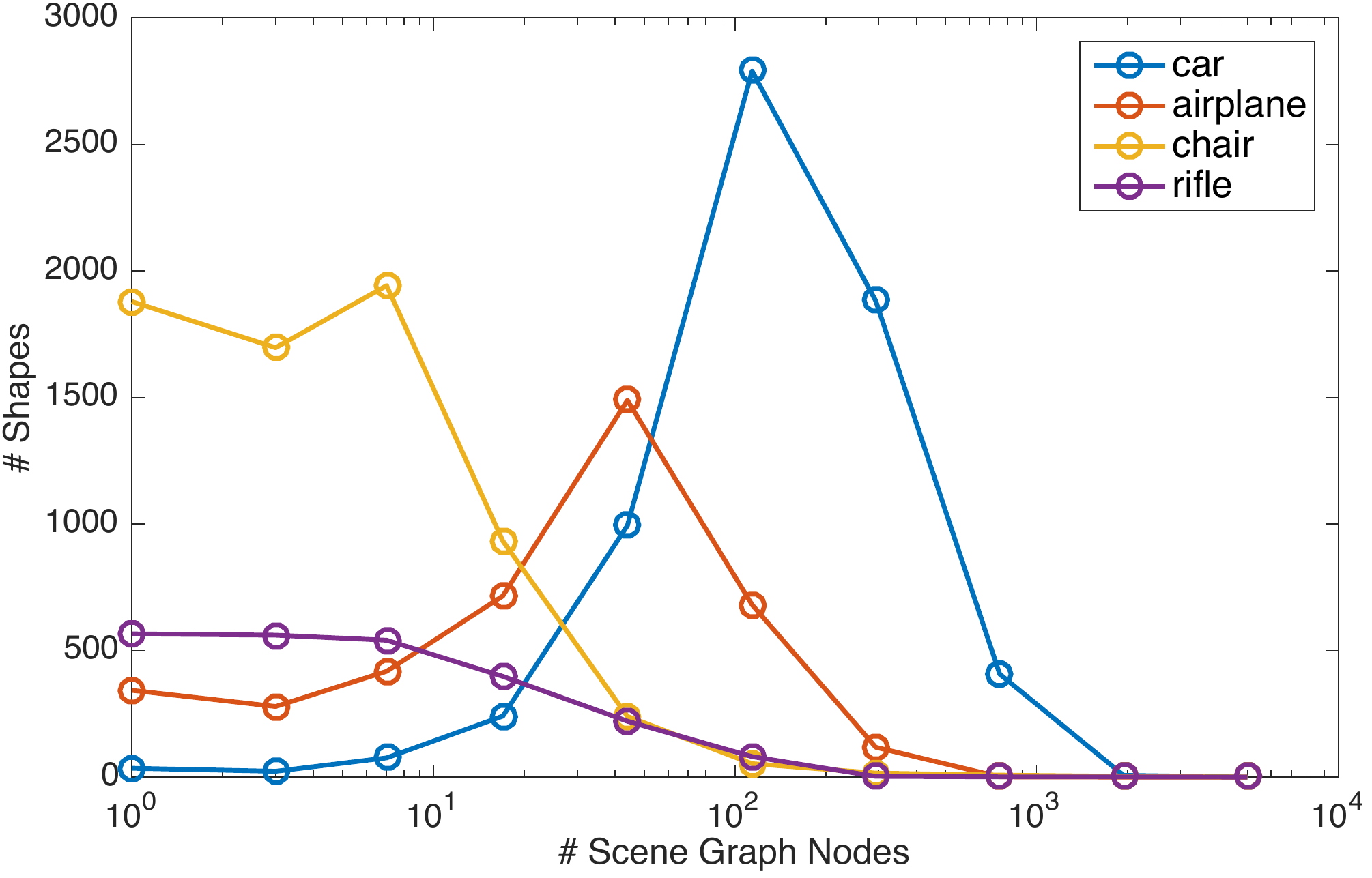}
  \vspace{-0.7cm}	
	\caption{Histogram of number of scene nodes for each shape in the raw datasets. 
	}
	  \vspace{-.3cm}	
	\label{fig:histogramnode}
\end{figure}

In this section we evaluate our method on ``in the wild'' data from public online repositories \edit{and on a standard benchmark}. We perform the evaluation by comparing with part-based analysis and segmentation techniques using novel metrics. 

\mypara{Input Data.}
We run our method on 9 shape categories from ShapeNetCore dataset \cite{chang2015shapenet}, a collection of 3D models obtained from various online repositories. We use this dataset for convenience, because the data has been preprocessed, cleaned, categorized, and put into common formats;
\edittwo{at present, it is the only known current dataset that satisfies our low-level preprocessing requirements.}
We excluded most categories ($\sim$40) because they only have a few hundred shapes or less, which is inadequate for our approach. We assume that a tag is sufficiently represented if it appears in at least 25 shapes, and we only analyze categories that have more than 2 such tags. Some categories have trivial geometry (e.g., mobile phones). Some categories do not provide enough parts with common labels (e.g., watercraft are very heterogeneous to the point of being disjoint sets of objects). 
The ShapeNetCore dataset currently contains a small subset of the available online repositories, which limits the data that we have immediately at hand. However, ShapeNetCore is growing; applying our method to much larger datasets is limited only by the nuisance of preprocessing heterogeneous datasets.
 
\begin{figure}[t!]
	\centering
	\includegraphics[width=1\columnwidth]{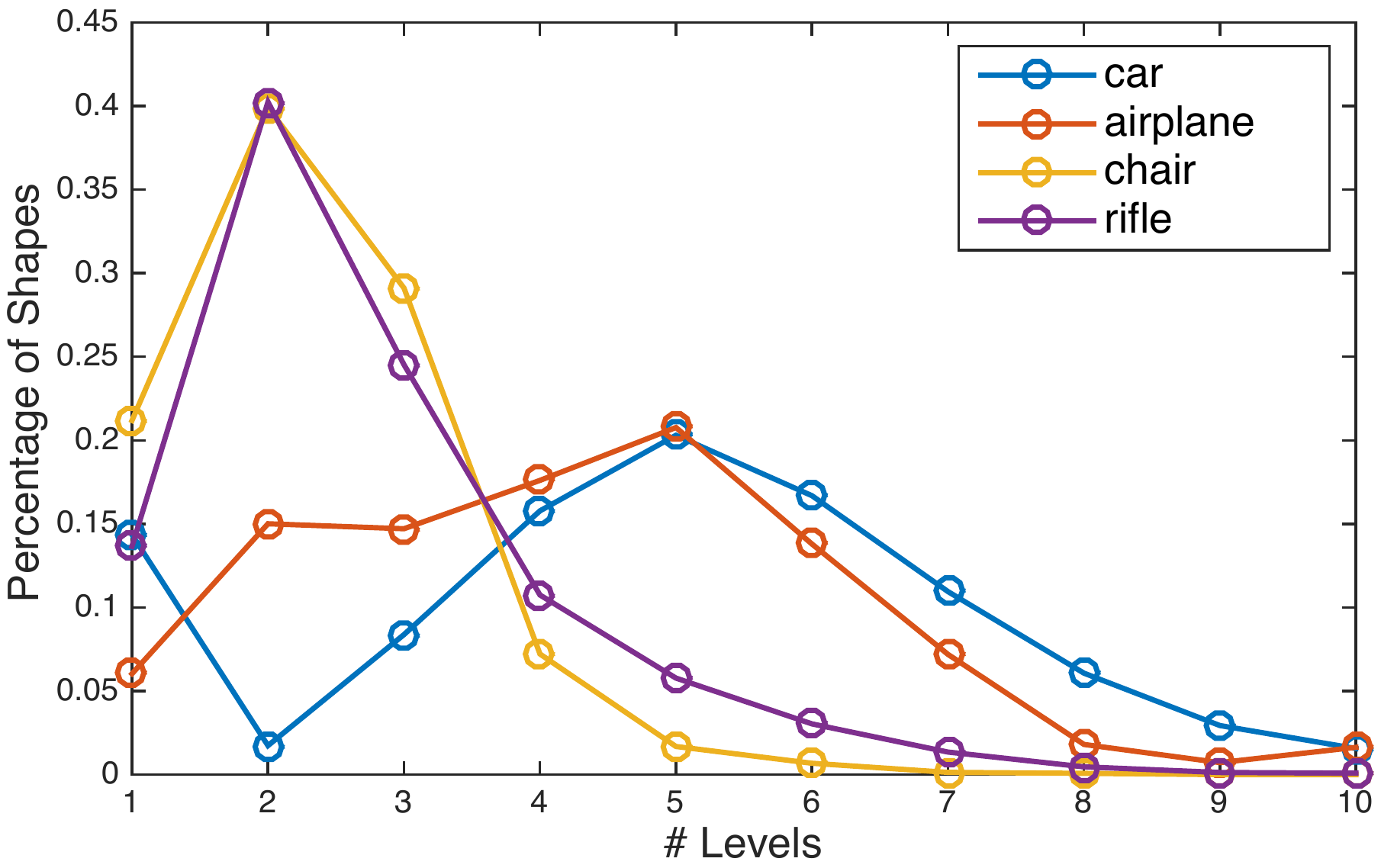}
  \vspace{-0.7cm}
	\caption{Histogram of number of levels in hierarchy for each shape in the raw datasets.\label{fig:histogramhierarchy} }
	  \vspace{-0.3cm}	
\end{figure}

\edit{Typical scene graphs in such online repositories are very diverse, including between one and thousands of nodes (Figure \ref{fig:histogramnode}), and ranging from flat to deep hierarchies (Figure \ref{fig:histogramhierarchy}).}
For each category, we also prescribe a list of relevant tags and possible synonyms. We automatically create a list of most-used tags for the entire category, and then manually pick relevant English nouns as the tag dictionary. 
Note that only a fraction of shapes have any parts with the chosen tags, and the frequency distribution over tag names is very uneven (Table \ref{tab:labelfrequency}, \textbf{Init} column). 

For a categories with $\sim$2000 shapes, the part-based analysis takes approximately one hour, and the segmentation training takes approximately 10 hours, each on a single Titan X GPU. \edit{Once trained, analysis of a new shape typically takes about 25 seconds, of which about 15 seconds is  extracting face features with non-optimized Matlab code.}

\begin{figure*}
	\centering
	\includegraphics[width=0.92\textwidth]{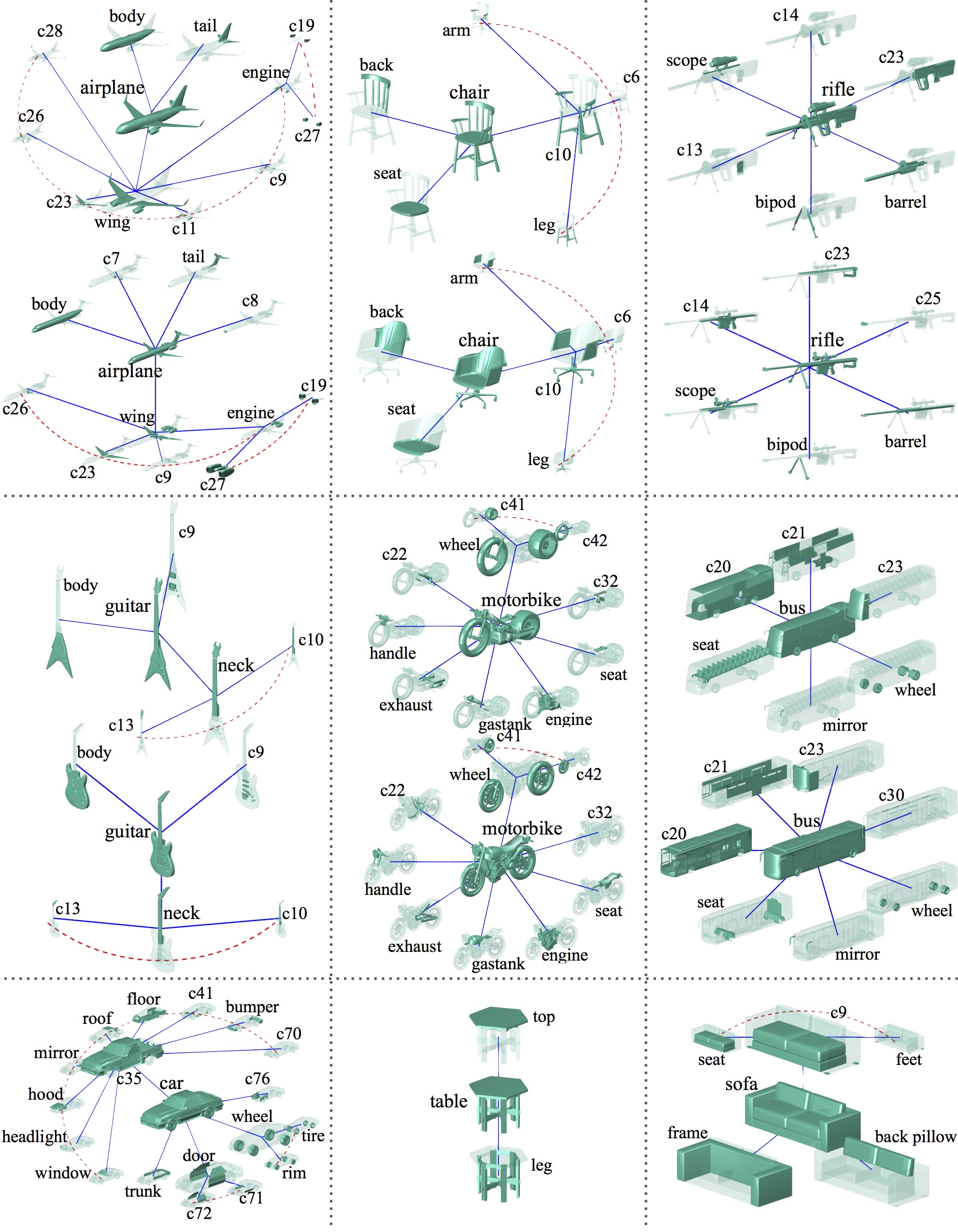}
	\caption{Hierarchical segmentation results. In each case, the input is a geometric shape. Our method automatically determines the segmentation into parts, the part labels and the hierarchy.}
	\label{fig:hierseg}
\end{figure*}

\mypara{Hierarchical Mesh Segmentation and Labeling.}
Figure~\ref{fig:hierseg} demonstrates some representative results produced by our 
hierarchical segmentation \edit{based on connected components} (Section~\ref{sec:seg}).  The resulting hierarchical segmentation vary in depth from flat (e.g., chairs) to deep (e.g., cars, airplanes), reflecting complexity of the corresponding object.  We also often extract  consistent part clusters, even if they do not have textual tags.
\edit{We found that analyzing shapes at the granularity of connected components is usually sufficient: the mean number of connected components per object in ShapeNet is 4169, and the largest connected component in shapes covers only $9.58\%$ of the total surface area on average: connected components tend to be small. These components are often aligned to part boundaries, for example, if one was to annotate ShapeNet Segmentation benchmark~\cite{yi2016scalable} by assigning a majority label to each connected component they would get $94\%$ of faces correct.} 

\begin{figure}
	\centering
	\includegraphics[width=0.5\textwidth]{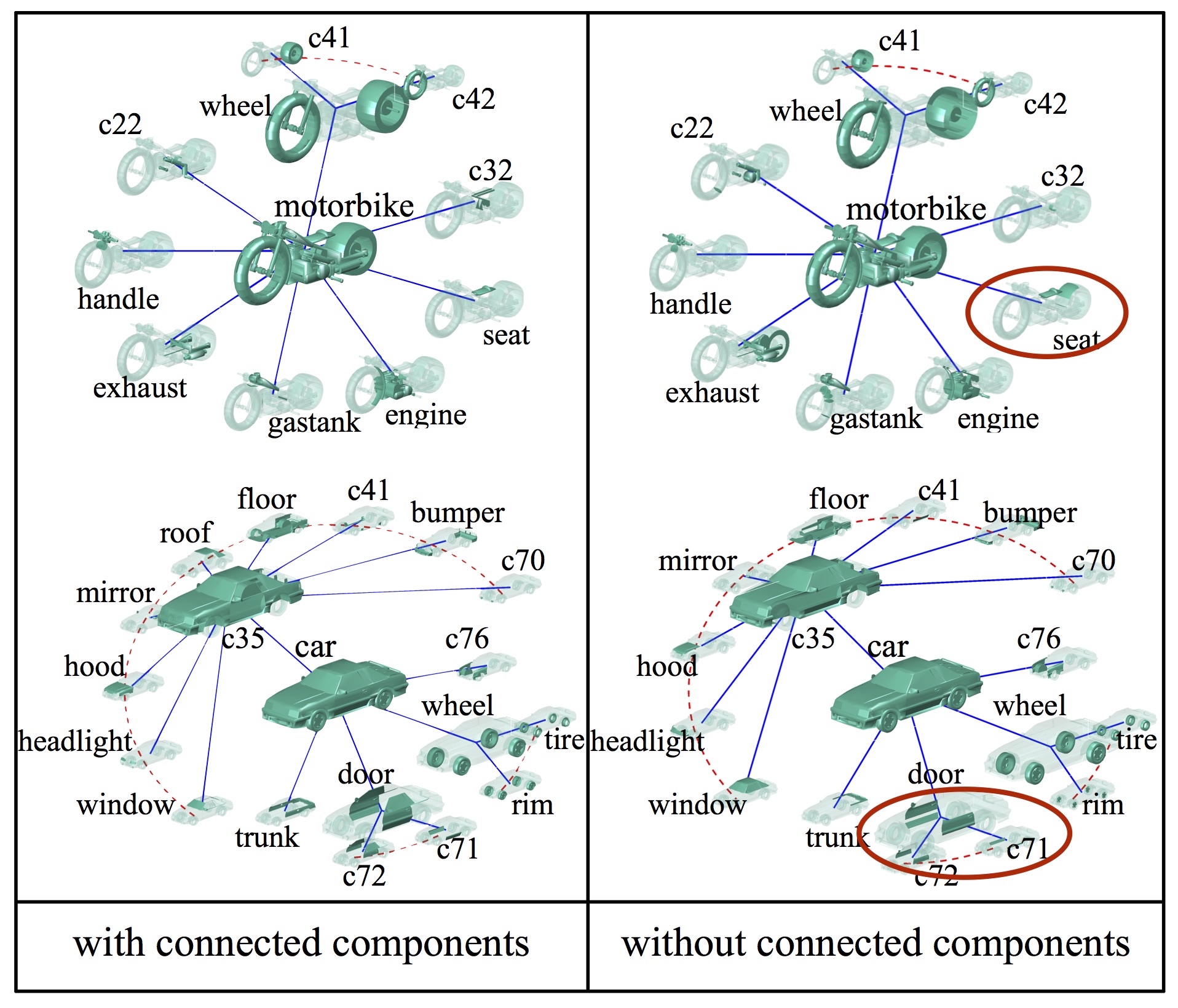}
  \vspace{-0.7cm}	
	\caption{Hierarchical segmentation results with and without the connected components assumption. Even without connected components, our method estimates complex hierarchical structures. For some models, the boundaries are less precise (see red higlights). We provide a full comparison to Figure~\ref{fig:hierseg} in supplemental material. }
  \vspace{-0.3cm}	
	\label{fig:hiersegnocc}
\end{figure}

\mypara{Segmentation without Connected Components.} 
%
In the case of applying per-face labeling, when connected components are not available, we observe similar results with this method as to those where the connected components are used. However, a few segments do not come out as cleanly-segmented on more complex models (see Figure~\ref{fig:hiersegnocc}).
Please refer to our supplementary material for qualitative results of this experiment.  
\edittwo{We tested our method on other datasets (Thingi10k \cite{thingi10k},  COSEG \cite{Wang12}), but were only able to test on a limited set of models, since only a few models in these datasets come from our training categories.}

\mypara{Tag prediction.} Table~\ref{tab:labelfrequency}, \emph{Final} column shows what fraction of training shapes received a particular tag after our part-based analysis (Section \ref{sec:method}). Note that an object may be missing a tag for several possible reasons: it could be misclassified, or because the object does not have that part, or does not have it segmented as a separate scene graph node. As evident from this quantitative analysis, the amount of training data we can use in subsequent analysis has drastically increased.  Please refer to supplementary material for visual examples from labeling results.

To evaluate tag prediction accuracy, we perform the following experiment. We hold out 30\% of the tagged parts during training, and evaluate labeling accuracy on these parts. As our method is based on nearest-neighbors (NN) classification, we compare against NN on features computed in the following ways:
(1) clustering with LFD, (2) clustering with $\bx$, (3) our method with no $E_c$ term, and  (4) No $E_m$ term. Results are reported in Table~\ref{tab:tagacc}. As shown in the table, our method significantly improves tag classification performance over the baselines. This experiment also demonstrates the value of our clustering and hierarchy terms $E_c$ and $E_m$.

%
%
%

\begin{table}[t!]
\small
\caption{\label{tab:labelfrequency} Tags and their frequency (percent of shapes that have a scene node (i.e., part) labeled with the corresponding tag) in the raw datasets and after our part based processing.}
  \vspace{-0.3cm}	
\centering
  \begin{tabular}{ @{}c@{}|@{}c@{}|@{}c@{}||@{}c@{}|@{}c@{}|@{}c@{}||@{}c@{}|@{}c@{}|@{}c@{} }
    \textbf{~Part~} & \textbf{~Init~} & \textbf{~Final~} & \textbf{~Part~}  & \textbf{~Init~} & \textbf{~Final~} & \textbf{~Part~}  & \textbf{~Init~} & \textbf{~Final~} \\ \hline
    \multicolumn{9}{c}{Category: \textbf{Car} (2287 shapes)} \\    \hline
    ~Wheel~  & ~17.4~ & ~96.4~ & ~Mirror ~& ~6.9 ~& ~68.2 ~ & ~Window~ &  ~9.5 ~ & ~71.8~\\  \hline
    ~Fender~  & ~1.1~ & ~40.4~ & ~Bumper ~& ~15.2 ~& ~63.0 ~ & ~Roof~ & ~2.4~ & ~48.2~ \\  \hline
    ~Exhaust~  & ~7.2~ & ~57.5~ & ~Floor~ & ~2.6~ & ~49.9~ & ~Trunk~ & ~4.5~ & ~60.6~ \\  \hline
    ~Door~ & ~19.9~ & ~67.8~ &  ~Spoiler~  & ~2.8~ & ~41.3~ & ~Rim~  & ~4.1~ & ~74.1~\\  \hline
    ~Headlight~ & ~14.6~ & ~61.9~ &  ~Hood~  & ~12.2~ & ~68.7~ & ~Tire~  & ~5.3~ & ~33.5~\\  \hline
    \multicolumn{9}{c}{Category: \textbf{Airplane} (2574 shapes)} \\    \hline
    ~Wing~  & ~5.9~ & ~86.9~ & ~Engine ~& ~5.5 ~& ~81.6 ~ & ~Body~ &  ~2.3 ~ & ~86.8~\\  \hline
    ~Tail~  & ~1.5~ & ~90.5~ & ~Missile~& ~0.4~& ~66.7~ & ~~ & ~~ & ~~ \\  \hline
    \multicolumn{9}{c}{Category: \textbf{Chair} (2401 shapes)} \\    \hline
    ~Arm~  & ~1.5~ & ~62.4~ & ~Leg ~& ~3.6 ~& ~71.0 ~ & ~Back~ &  ~0.9 ~ & ~50.8~\\  \hline
    ~Seat~  & ~2.4~ & ~83.1~ & ~Wheel~& ~1.4~& ~34.6~ & ~~ & ~~ & ~~ \\  \hline
    \multicolumn{9}{c}{Category: \textbf{Table} (2355 shapes)} \\    \hline
    ~Top~  & ~1.9~ & ~81.5~ & ~Leg ~& ~6.4 ~& ~85.8 ~ & ~~ &  ~ ~ & ~~\\  \hline
    \multicolumn{9}{c}{Category: \textbf{Sofa} (1243 shapes)} \\    \hline
    ~BackPillow~  & ~4.7~ & ~79.3~ & ~Seat ~& ~2.7 ~& ~63.0 ~ & ~Feet~ &  ~ 3.1~ & ~41.5~\\  \hline
    \multicolumn{9}{c}{Category: \textbf{Rifle} (994 shapes)} \\    \hline
    ~Barrel~  & ~1.2~ & ~62.9~ & ~Bipod ~& ~1.7 ~& ~47.6 ~ & ~Scope~ &  ~3.2 ~ & ~66.7~\\  \hline
    ~Stock~  & ~3.2~ & ~56.0~ & ~~& ~~& ~~ & ~~ & ~~ & ~~ \\  \hline
    \multicolumn{9}{c}{Category: \textbf{Bus} (713 shapes)} \\    \hline
    ~Seat~  & ~4.8~ & ~47.0~ & ~Wheel ~& ~8.3 ~& ~88.9 ~ & ~Mirror~ &  ~1.7 ~ & ~42.4~\\  \hline
    \multicolumn{9}{c}{Category: \textbf{Guitar} (491 shapes)} \\    \hline
    ~Neck~  & ~3.1~ & ~75.8~ & ~Body ~& ~0.8 ~& ~67.6 ~ &  &    & \\  \hline
    \multicolumn{9}{c}{Category: \textbf{Motorbike} (281 shapes)} \\    \hline
    ~Seat~  & ~23.5~ & ~49.2~ & ~Engine ~& ~21.7 ~& ~84.0 ~ & ~Gastank~ &  ~14.6 ~ & ~73.7~\\  \hline
    ~Exhaust~  & ~1.8~ & ~71.5~ & ~Handle~& ~8.9~& ~75.8~ & ~Wheel~ & ~41.6~ & ~98.9~ \\  \hline
  \end{tabular}
  \vspace{-0.3cm}	  
\end{table}

\begin{figure}[t!]
	\centering
	\includegraphics[width=1\columnwidth]{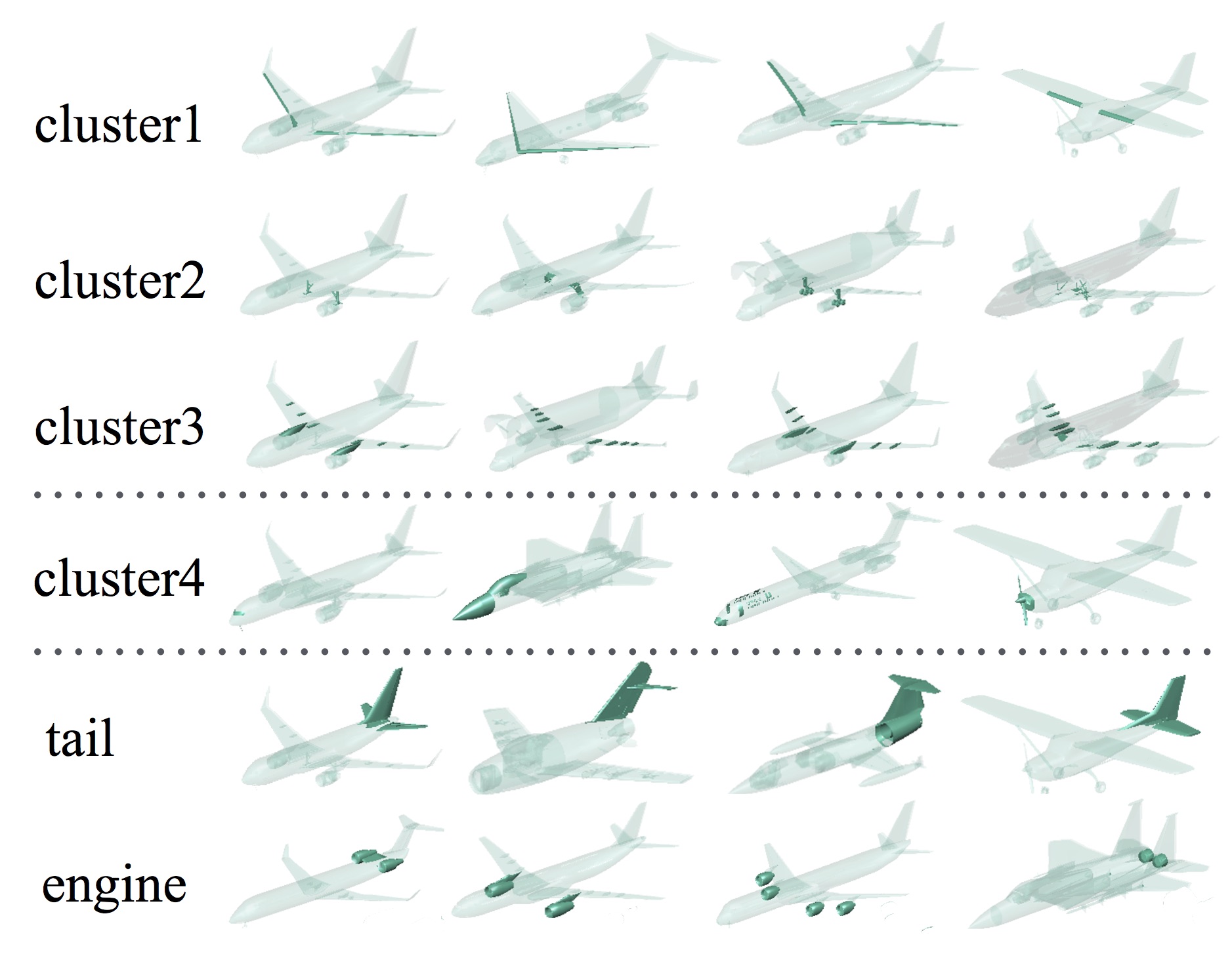}
  \vspace{-0.7cm}	
	\caption{Visualization of typical clusters. Note that some clusters have labels that were propagated from the tags, whereas some have generic labels indicating that they were discovered without any tag information. }
  \vspace{-0.3cm}	
	\label{fig:clustervis}
\end{figure}

\mypara{Cluster evaluation.}
Figure~\ref{fig:clustervis} (bottom) demonstrates some parts grouped by our method in the part-based analysis (Section \ref{sec:method}). We also note that some clusters combine unrelated parts, and we believe that they serve as null clusters for outliers. 

As we do not have ground truth for the unlabeled clusters, we instead evaluate the ability of our learned embedding to cluster parts, with the user-provided labels can serve as ground truth. We split the dataset for a category into \textit{training tags} and \textit{test tags}. We run the part-based analysis on all shapes, but provide only the training subset of tags to the algorithm.  This gives an embedding $f$, and we can evaluate how  good $f$ is for clustering. This is done by running $k$-means clustering on the parts that correspond to the test tags, with $k$ set to the number of test tags. The clustering result is then compared to the test tag labeling by normalized Mutual Information.  This process is repeated in a 3-fold cross-validation. The baseline scores are especially low on categories with few parts, like Bus and Table. 
Table~\ref{tab:embedeva} shows quantitative results; our method performs significantly better than the baselines, including using $k$-means on Light Field Descriptors (LFD), and omitting the clustering term $E_c$ from the objective.

%
%
%

%

\begin{figure}[t!]
	\centering
	\includegraphics[width=1\columnwidth]{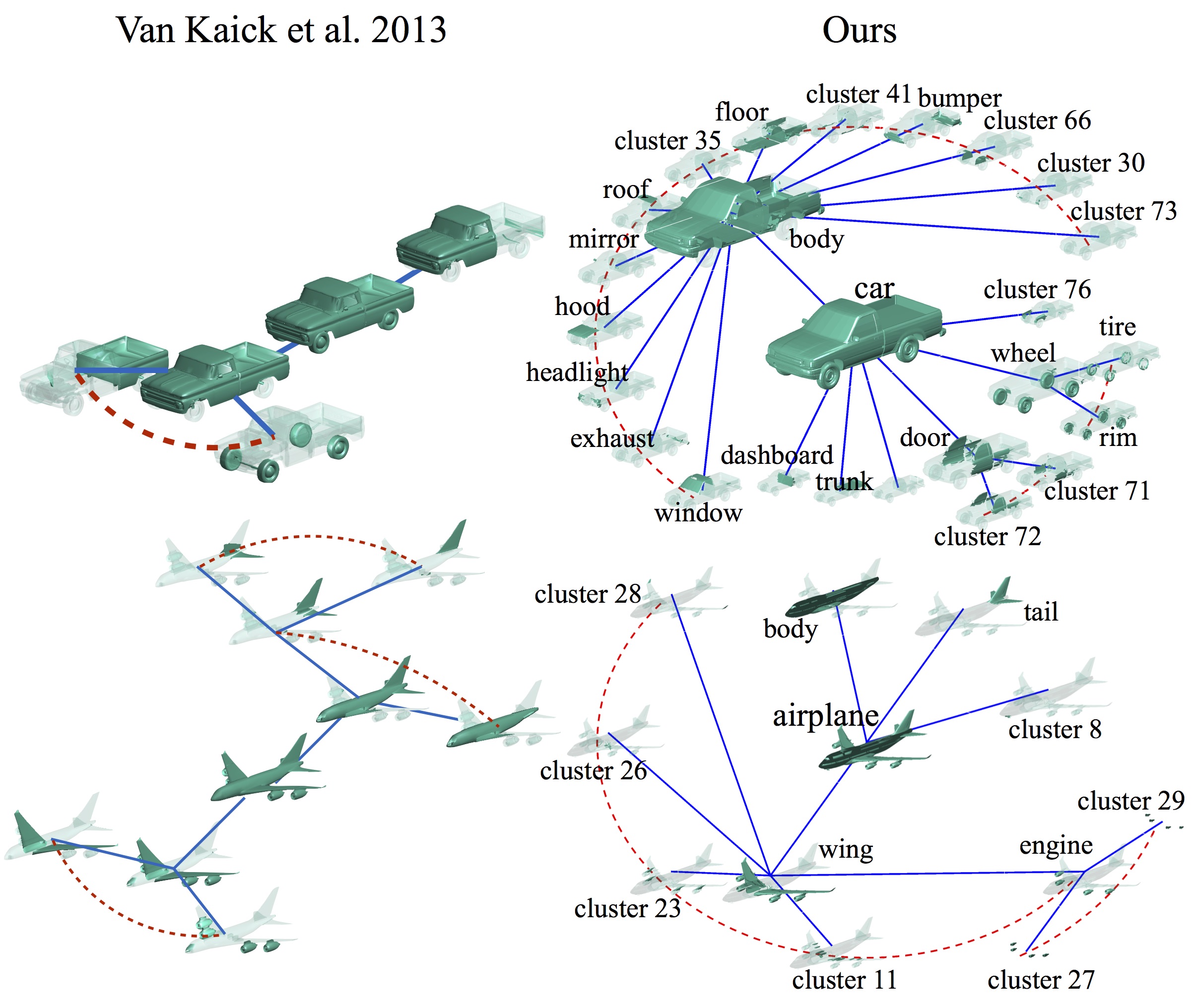}
  \vspace{-0.7cm}	
	\caption{Comparison with \protect\cite{van2013co}. We show a hierarchy produced by their approach (left) and a hierarchy of the most similar model in our database (right). Our hierarchies have labels and provide finer details. }
  \vspace{-0.4cm}	
	\label{fig:vanKaick}
\end{figure}

\mypara{Comparison to Unsupervised Co-Hierarchy.}
Van Kaick et al.\ \shortcite{van2013co} propose an unsupervised approach for establishing consistent hierarchies within an object category. Their method was developed for small shape collections and requires hours of computation for 20 models, which makes it unsuitable for ShapeNet data. On the other hand, since we assume that some segments have textual tags, we also cannot run our method on their data. Given these constraints, we show a qualitative comparison to their method. In particular, we picked the first car and first airplane in their dataset, and retrieved the most similar models in ShapeNet using lightfield descriptors. Figure~\ref{fig:vanKaick} demonstrates their and our hierarchies side-by-side. Note that our method generates more detailed hierarchies and also provides textual tags for parts.

\mypara{Comparison to Supervised Segmentation.}
Since there are no large-scale hierarchical segmentation benchmarks, we test our method on the segmentation dataset provided by Yi et al.~\shortcite{yi2016scalable}. We emphasize that the benchmark contains much coarser segmentations than those we can produce, and does not include hierarchies. We take the intersection of our 9 categories and the benchmark, which yields the following six categories for quantitative evaluation: car, airplane, motorbike, guitar, chair, table. 

\edit{
Since other techniques do not leverage connected components, we evaluate per-face classification from unary terms only, comparing the per-face classification prediction (Eq.~\ref{eq:segunary}) to results from Yi et al.~\shortcite{yi2016scalable} trained only on benchmark data.}

Our training data is sampled from a different data distribution than the benchmark; repurposing a model from one training set to another is a problem known as \textit{domain adaptation.} The first approach we test is to directly map the labels predicted by our classifier to benchmark labels. \edit{The second approach is to obtain 5 training examples from the benchmark, and train a Support Vector Machine classifier to predict benchmark labels from our learned features $\{ g(\bx)\}$ (Sec.~\ref{sec:method}). The resulting classifier is the softmax of $\{\eta_i g(\bx)\}$, where $\eta_i$ are the SVM parameters for $i^{th}$ label. As baseline features, we also test $k$-means clustering with LFD features over all input parts, where $k$ is the same as the number of clusters used by our method.} 

\begin{figure}[t!]
	\centering
	\includegraphics[width=1\columnwidth]{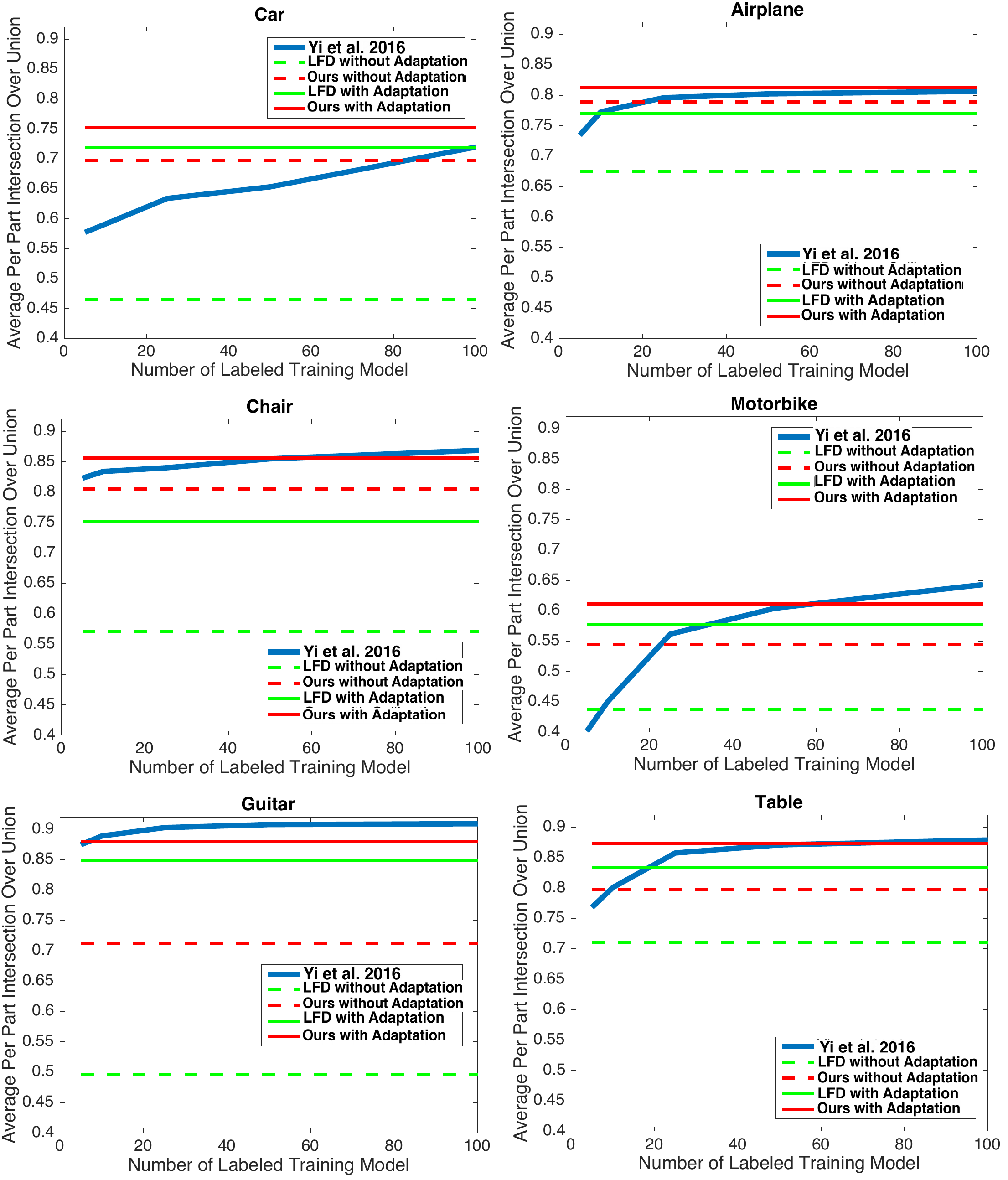}
  \vspace{-0.5cm}	
	\caption{Comparison with \protect\cite{yi2016scalable}.  Segmentation accuracy scores are shown; higher is better. The blue curves show the results of Yi et al.~as a function of training set size. The red dashed lines show the result of our method without applying domain adaption, and the red solid lines show our method with domain adaptation by training on 5 benchmark models. For the more complex classes, Yi et al.'s method requires 50-100 training meshes to match our performance with only 5 benchmark training meshes.
	}
  \vspace{-0.3cm}		
	\label{fig:segeva}
\end{figure}

\edit{Results of supervised segmentation comparison experiments are shown in Figure~\ref{fig:segeva}.} Without training on our features, the method of Yi et al.~\shortcite{yi2016scalable} requires 50-100 benchmark training examples in order to match the results we get with only 5 benchmark examples. Although our method is trained on many ShapeNet meshes, these meshes did not require any manual labeling. This illustrates how our method, trained on freely-available data, can be cheaply adapted to a new task.

\begin{figure}[t!]
	\centering
	\includegraphics[width=1\columnwidth]{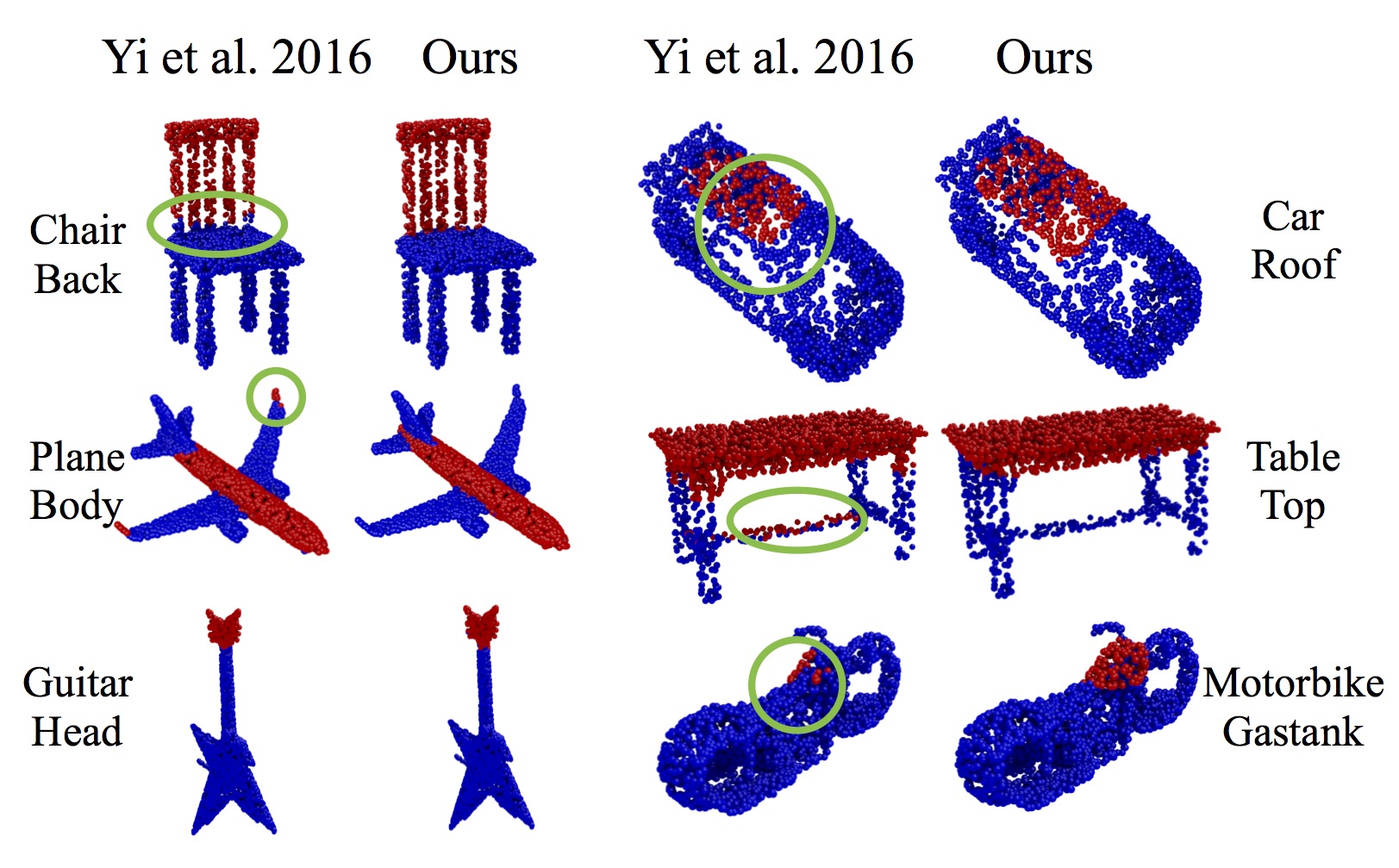}
  \vspace{-0.5cm}	
	\caption{Qualitative comparison with \protect\cite{yi2016scalable}. For a fair comparison, we use 10 models for training in \protect\cite{yi2016scalable} and we use the same 10 models for domain adaptation in our approach.}
  \vspace{-0.3cm}	
	\label{fig:segcomp}
\end{figure}

\edit{Figure~\ref{fig:segcomp} shows qualitative results from comparison with Yi et al.~\shortcite{yi2016scalable}, where we use 10 models for training in \protect\cite{yi2016scalable} followed by the domain adaptation through using the same 10 models in our approach. }

\begin{table*}[t!]
\caption{\label{tab:embedeva} Clustering evaluation with different embedding/features. Scores shown are Normalized Mutual Information between the estimated clusters and the user-provided tags. A perfect score corresponds to NMI of 1. Note that the user-provided tags may themselves be noisy, so perfect scores are very unlikely.
}
  \vspace{-0.2cm}
\centering
  \begin{tabular}{ | l | c || c | c | c | c | c | c | c | c | c |}
    \hline
    Category & Mean & Car & Airplane & Chair & Table & Motorbike & Bus & Guitar & Rifle & Sofa \\ \hline
    Chance & 0.034 & 0.019 & 0.010 & 0.040 & 0.005 & 0.020 & 0.026 & 0.018 & 0.176 & 0.032\\ \hline
    LFD & 0.336 & 0.521 & 0.315 & 0.350 & 0.238 & 0.576 & 0.292 & 0.034 & 0.379 & 0.297\\ \hline
    $\bx$ (part features - App.\protect~\ref{appx:features}) & 0.348 & 0.551 & 0.264 & 0.352 & 0.238 & 0.607 & 0.313 & 0.101 & 0.405 & 0.297\\ \hline
    No $E_c$ term & 0.406 & 0.626 & 0.377 & 0.346 & 0.124 & 0.564 & 0.260 & 0.498 & 0.445 & 0.408\\ \hline
    No $E_m$ term & 0.561 & 0.695 & 0.568 & {\bf 0.622} & 0.445 & 0.659 & 0.367 & 0.655 & 0.514 & 0.566\\ \hline
    Ours & {\bf 0.573} & {\bf 0.712} & {\bf 0.575} & 0.619 & {\bf 0.448} & {\bf 0.678} & {\bf 0.371} & {\bf 0.655} & {\bf 0.526} & {\bf 0.571}\\ \hline
  \end{tabular}
\end{table*}

\begin{table*}[t!]
\caption{\label{tab:tagacc} Part tagging accuracy comparison. We hold out tags for $30\%$ originally tagged parts in the input data, and report testing accuracy on the held out set.
}
  \vspace{-0.2cm}
\centering
  \begin{tabular}{ | l | c || c | c | c | c | c | c | c | c | c |}
    \hline
    Category & Mean & Car & Airplane & Chair & Table & Motorbike & Bus & Guitar & Rifle & Sofa \\ \hline
    Chance & 0.139 & 0.044 & 0.136 & 0.172 & 0.148 & 0.149 & 0.100 & 0.252 & 0.092 & 0.162\\ \hline
    LFD & 0.790 & 0.530 & 0.823 & 0.775 & 0.745 & 0.829 & 0.813 & 0.976 & 0.723 & 0.892\\ \hline
    $\bx$ (part features - App.\protect~\ref{appx:features})  & 0.823 & 0.584 & 0.832 & 0.812 & 0.772 & 0.874 & 0.822 & 0.976 & 0.818 & 0.920\\ \hline
    No $E_c$ term & 0.840 & 0.694 & 0.821 & 0.749 & 0.910 & 0.860 & 0.911 & 0.982 & 0.772 & 0.864\\ \hline
    No $E_m$ term & 0.899 & 0.701 & 0.934 & 0.902 & {\bf 0.926} & 0.865 & 0.884 & 0.991 & 0.936 & 0.953\\ \hline
    Ours & {\bf 0.910} & {\bf 0.709} & {\bf 0.97} & {\bf 0.905} & 0.921 & {\bf 0.878} & {\bf 0.884} & {\bf 0.994} & {\bf 0.951} & {\bf 0.979}\\ \hline
  \end{tabular}
\end{table*}


\section{Discussions and Conclusion}

We have proposed a novel method for mining consistent hierarchical shape models from massive but sparsely annotated scene graphs ``in the wild.'' As we analyze the input data, we jointly embed parts to a low-dimensional feature space, cluster corresponding parts, and build a probabilistic model for hierarchical relationships among them. We demonstrated that our model can facilitate hierarchical mesh segmentation and were able to extract complex hierarchies and identify small segments in 3D models from various shape categories. Our method can also provide a valuable boost for supervised segmentation algorithms.
The goal of our current framework is to extract as much structure as possible from raw noisy, sparsely tagged scene graphs that exist in online repositories.
In the future, we believe that using such freely-available information will provide enormous opportunities for shape analysis. 


\edit{Developing Convolutional Neural Networks for surfaces is a very active area right now, e.g., \cite{Guo:2015}. Our segmentation training loss functions are largely agnostic to the model representation, and it ought to be straightforward to train a ConvNet on our training loss, for any ConvNet that handles disconnected components.}

\edit{
Though effective as evidenced by experimental evaluations, several issues are not completely addressed yet. Our model currently relies on heuristic selection of the number of clusters $k$, and this could be chosen automatically. We could also relax the assumption that each part with a given label may have only one possible parent label, to allow more general shape grammars \cite{talton2012learning}.}

\edit{
Our method has obtained about $13$K 3D training models with roughly consistent segmentation, but these have not been human-verified. We also believe that our approach could be leveraged together with crowdsourcing techniques \cite{yi2016scalable} to efficiently yield very large, detailed, segmented, and verified shape databases.}

It would also be interesting to explore how well the information learned from one object category may transfer to other object categories. For example, ``wheel'' can be found in ``cars'' and ``motorbikes'', sharing similar geometry and sub-structures. The observation provides the opportunity for not only the transfer of part embeddings but also the part relationships. With the growth of online model repositories, such transfer learning ability would be more important and relevant towards more efficient expanding of our current dataset.


\bibliographystyle{acmsiggraph}
\nocite{*}
\bibliography{template}

\appendix

\appendix 

\begin{figure}
	\centering
	\includegraphics[width=1\columnwidth]{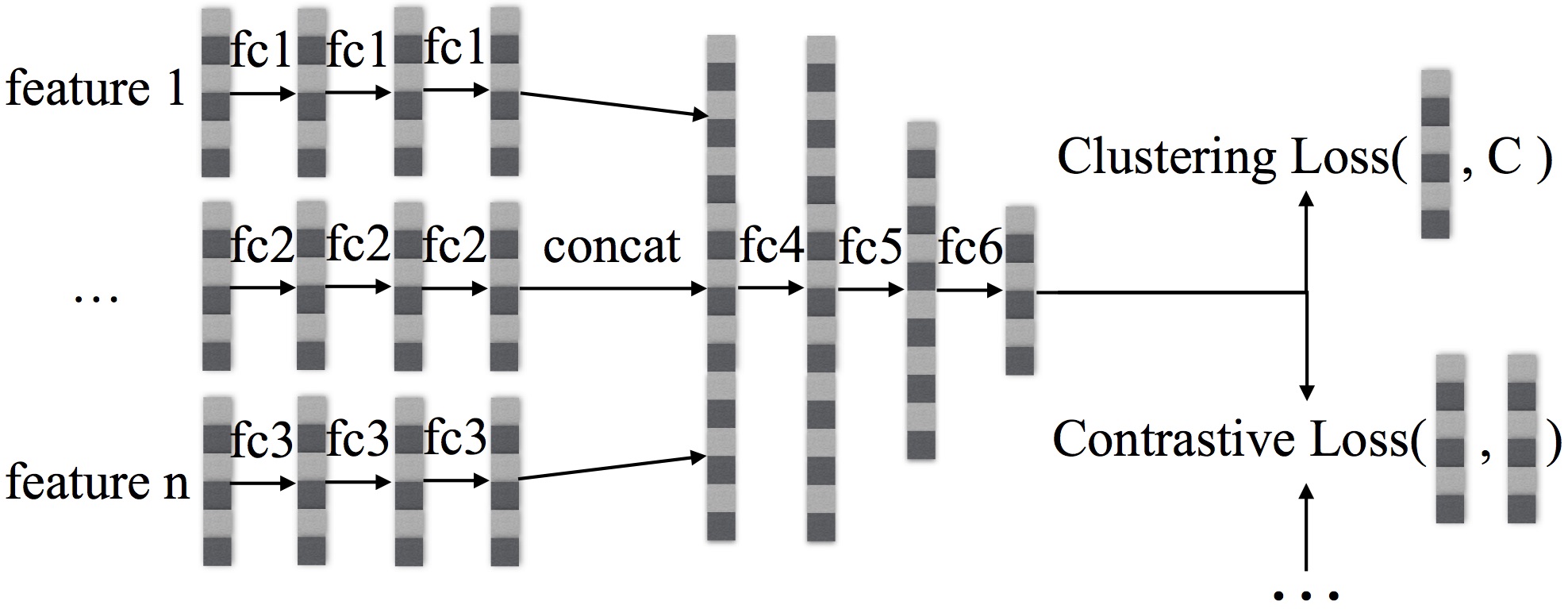}
\vspace{-0.7cm}		
	\caption{Embedding network $f$ architecture for parts. Different input part features go through a stack of fully connected layers and then are concatenated and go through additional fully connect layers to generate output of $f$. The contrastive loss also takes output of another part, from an identical embedding branch as input, which is omitted in this figure for brevity. 
	}
\vspace{-0.3cm}		
	\label{fig:network_arc}
\end{figure}

\section{Part Features and Embedding Network}
\label{appx:features}
We compute per-part geometric features which are further used for joint part embedding and clustering (Section~\ref{sec:method}). The feature vector $\bx_{ij}$ includes 3-view lightfield descriptor~\cite{Chen2003lightfield} (with HOG features for each view), center-of-mass, bounding box diameter, approximate surface area (fraction of voxels occupied in 30x30x30 object grid), and local frame in PCA coordinate system (represented by $3\times3$ matrix $M$). To mitigate reflection ambiguities for local frame we constraint all frame axes to have positive dot product with $z$-axis (typically up) of the global frame.  For lightfield descriptor we normalize the part to be centered at origin and have bounding box diameter 1, for all other descriptors we normalize the mesh in the same way. We mitigate reflection ambiguities by constraining all frame axes to have positive dot product with the $z$-axis of the global frame.
The neural network embedding $f$ is visualized in Figure \ref{fig:network_arc}, and,
in Table~\ref{tab:network_param}, we show the embedding network parameters, where we alter first few fully connected layers to allocate more neurons for richer features such as LFD.

\begin{table}
\caption{\label{tab:network_param} Embedding network $f$ output dimensionalities after each layer.}
\vspace{-0.3cm}		
\centering
  \begin{tabular}{ | l | c | c | c | c | c | c | c |}
    \hline
    feature & fc1 & fc2 & fc3 & concat & fc4 & fc5 & fc6 \\\hline
    LFD  & 128 & 256  & 256 & \multirow{5}{*}{512} & \multirow{5}{*}{256} & \multirow{5}{*}{128} & \multirow{5}{*}{64}\\\cline{1-4}
    PCA Frame & 16 & 32  & 64 &&&& \\ \cline{1-4}
    CoM & 16 & 64  & 64 &&&& \\ \cline{1-4}
    Diameter & 8 & 32  & 64 &&&& \\ \cline{1-4}
    Area & 8 & 32  & 64 &&&& \\ \hline
  \end{tabular}
\end{table}

\section{Face Features and Classifier Network}
\label{appx:featuresfaces}
We compute per-face geometric features $\by$ which are further used for hierarchical mesh segmentation (Section~\ref{sec:seg}). These features include spin images (SI)~\cite{Johnson1999spinimages}, shape context (SC)~\cite{Belongie2002shapecontext}, distance distribution (DD)~\cite{osada2002SD}, local PCA (LPCA) (where $\lambda_i$ are eigenvalues of local coordinate system, and features are $\lambda_1/\sum\!\lambda_i, \lambda_2/\sum\!\lambda_i, \lambda_3/\sum\!\lambda_i, \lambda_2/\lambda_1, \lambda_3/\lambda_1, \lambda_3/\lambda_2$), local point position variance (LVar), curvature, point position (PP) and normal (PN). To compute local radius for the feature computation we sample 10000 points on the entire shape and use 50 nearest neighbors. We use the same architecture as part embedding network $f$ (Fig.~\ref{fig:network_arc}) for face classification, but with different loss function (Eq.~\ref{eq:robustsoftmax}) and network parameters, which are summarized in Table~\ref{tab:facenet_param}.

\begin{table}
\caption{\label{tab:facenet_param} Face classification network parameters.}
\vspace{-0.3cm}		
\centering
  \begin{tabular}{ | l | c | c | c | c | c | c | c |}
    \hline
    feature & fc1 & fc2 & fc3 & concat & fc4 & fc5 & fc6 \\\hline
    Curvature  & 32 & 64  & 64 & \multirow{8}{*}{640} & \multirow{8}{*}{256} & \multirow{8}{*}{128} & \multirow{8}{*}{128}\\\cline{1-4}
    LPCA & 64 & 64  & 64 &&&& \\ \cline{1-4}
    LVar & 32 & 64  & 64 &&&& \\ \cline{1-4}
    SI & 128 & 128  & 128 &&&& \\ \cline{1-4}
    SC & 128 & 128  & 128 &&&& \\ \cline{1-4}
    DD & 32 & 64  & 64 &&&& \\ \cline{1-4}
    PP & 16 & 32  & 64 &&&& \\ \cline{1-4}
    PN & 16 & 32  & 64 &&&& \\ \hline
  \end{tabular}
\end{table}

\section{E-step Update}
\label{app:pupdate}

\edit{In the E-step, the assignment probabilities are iteratively updated.
For each node $(i,j)$, the probability that it is assigned to label $k$ is updated as:
\begin{align}
p*_{ijk} &\leftarrow
\exp\left( \lambda_m \sum_{a\in C(i,j), \ell}
p_{ia\ell} \ln \bM_{k\ell} +
\lambda_m \sum_{b= P(i,j), \ell} p_{ib\ell} \ln \bM_{\ell k}
 \right. \nonumber\\
&\qquad\qquad\left.-  \lambda_c ||f(\bx_{ij} -\bc_{k}||_1\right) \\
p_{ijk} &\leftarrow \frac{p*_{ijk}}{\sum_\ell p^*_{ij\ell}}
\end{align}
where  $C(i,j)$ is set of children of node $(i,j)$ and $P(i,j)$ is the parent node.
A joint closed-form update to all assignments could be computed using Belief Propagation, but we did not try this.
}

\end{document}